\documentclass{article}
\usepackage{graphicx}
\usepackage{xcolor}
\usepackage{amsmath, amsthm, amssymb,mathtools}
\usepackage{times}

\newcommand{\bmp}{\begin{minipage}}
\newcommand{\emp}{\end{minipage}}

\newcommand{\M}{\mbox{\it MinMax}}
\newcommand{\PK}{\mathcal P}
\newcommand{\kM}{\mbox{\it k}}
\newcommand{\oM}{\mbox{\it 1}}

\newcommand{\n}{[n]}
\newcommand{\lrf}[1]{\xleftrightarrow{#1}}
\newcommand{\rf}[1]{\xrightharpoondown{#1}}
\newcommand{\lf}[1]{\xleftharpoondown{#1}}
\newcommand{\Sil}{{\it SilNB}}

\newtheorem{thm}{Theorem}
\newtheorem{cor}{Corollary}
\newtheorem{fait}{Claim}
\newtheorem{rmk}{Remark}
\newtheorem{defin}{Definition}
\newtheorem{ex}{Example}
\newtheorem{pb}{Problem}

\newcommand{\bp}{\begin{pb}\rm}
\newcommand{\ep}{\end{pb}}
\newcommand{\br}{\begin{rmk}\rm}
\newcommand{\er}{\end{rmk}}
\newcommand{\bdefin}{\begin{defin}\rm}
\newcommand{\edefin}{\end{defin} }
\newcommand{\bex}{\begin{ex}\rm}
\newcommand{\eex}{\end{ex}}
\newcommand{\bthm}{\begin{thm}}
\newcommand{\ethm}{\end{thm}}
\newcommand{\bcor}{\begin{cor}}
\newcommand{\ecor}{\end{cor}}
\newcommand{\bfn}{\begin{fait}}
\newcommand{\efn}{\end{fait}}

\usepackage{algorithm}
\usepackage{algorithmic}



\renewcommand{\Box}{\rule{1.5mm}{3mm}}

\begin{document}

%\begin{frontmatter}
%\title{Common Intervals, Nested Common Intervals and Conserved Intervals of $K$ Permutations: a Common Approach}

\begin{center}
{\bf\large Permutation Reconstruction from $\M$-Betweenness Constraints}\\

%{\bf\large Finding Common Intervals, Nested Common Intervals and Conserved\\
%\vspace*{0.2cm}
%
% Intervals of $K$ Permutations: a Common Approach}\\

%\author{Irena Rusu\fnref{label2}}
\vspace*{1cm}

Irena Rusu\footnote{Irena.Rusu@univ-nantes.fr}

L.I.N.A., UMR 6241, Universit\'e de Nantes, 2 rue de
la Houssini\` ere,\\

 BP 92208, 44322 Nantes, France
\end{center}

%\fntext[label2]{Corresponding author. Phone: 033.2.51.12.58.16 Fax: 033.2.51.12.58.12 Email: Irena.Rusu@univ-nantes.fr}

%\address{L.I.N.A., UMR 6241, Universit\'e de Nantes, 2 rue de
%la Houssinière, BP 92208, 44322 Nantes, France}
%\begin{abstract}
\vspace*{1cm}

\hrule
\vspace{0.3cm}

\noindent{\bf Abstract}

In this paper, we investigate the  reconstruction of permutations on $\{1, 2, \ldots, n\}$ from
betweenness constraints involving the minimum and the maximum element located between 
 $t$ and $t+1$, for all $t=1, 2, \ldots, n-1$.  We propose two
variants of the problem (directed and undirected), and focus first on the directed version,
for which we draw up general features and design a polynomial algorithm 
in a particular case. Then, we investigate
necessary and  sufficient conditions for the uniqueness of the reconstruction
in both directed and undirected versions, using a parameter $k$ whose variation controls the
stringency of the betweenness constraints. We finally point out open problems.
\medskip

\vspace{0.2cm}

\hrule

\section{Introduction}

The {\sc Betweenness} problem is motivated by physical mapping in molecular biology and the design of circuits 
\cite{opatrny1979total}.
In this problem, we are given the set $\n :=\{1, 2, \ldots, n\}$, for some positive integer $n$,  
and a set of $m$ {\em betweenness constraints} ($m>0$), each represented as a triple $x\lrf{a}y$ with $x,a,y\in\n$ and signifying that $a$ is required to be between $x$ and $y$. The goal is to find a 
permutation on $\n$ satisfying a maximum number of betweenness constraints. In \cite{opatrny1979total},
it is shown that the {\sc Betweenness} problem is NP-complete even in the particular case where all the
constraints have to be satisfied.

In this paper we are interested in a problem related to the {\sc Betweenness} problem,  which also
finds its motivations in molecular biology. Given $K$ ($K\geq 2$) permutations on the
same set $\n$, representing $K$ genomes given by the sequences of their genes, a {\em common interval} 
of these permutations is a subset of $\n$ whose elements are consecutive ({\em i.e.} they form an interval) 
on each of the $K$ permutations. Common intervals thus represent regions of the genomes which have identical 
gene content, but possibly different gene order. Computing common intervals or specific subclasses of
them in linear time (up to the number of output intervals) has been done by case-by-case approaches
until recently, when we  proposed \cite{IR} a common linear framework, whose basis is the notion of
$\M$-profile. The $\M$-profile of a permutation $P$ forgets the order of the elements in a permutation, 
and keeps only essential betweenness information, defined as, for each $t\in [n-1]$, the minimum
and maximum value in the interval delimited by the elements $t$ (included) and $t+1$ (included) 
on $P$ (with no restriction on the relative positions of $t$ and $t+1$ on $P$). When $K$ permutations 
are available, their $\M$-profile is defined similarly, by considering for every  $t\in [n-1]$ the global 
minimum and the global maximum of the $K$ intervals delimited by $t$ and $t+1$ on the $K$ permutations. 
We show in \cite{IR} that, assuming the permutations have been renumbered such that one of them is
the identity permutation, the $\M$-profile of $K$ permutations is all we need to find common intervals,
as well as all the specific subclasses of common intervals defined in the literature, in linear time (up to the number of output intervals).

Hence, the $\M$-profile is a simplified representation of a (set of) permutation(s), which is sufficient
to efficiently solve a number of problems related to finding common intervals in permutations. 
Moreover, it may be computed in linear time \cite{IR}. However, it can be easily seen that distinct
(sets of) permutations may have the same $\M$-profile, implying that the $\M$-profile captures a part,
but not all, of the information in the (set of) permutation(s). 

In this paper, we study the reconstruction of a permutation from a given $\M$-profile, and discuss possible
generalizations.

\section{Definitions and Problems}\label{sect:def}

In the remaining of the paper, permutations are defined on $\n$ and are increased with elements $0$ and $n+1$,
added respectively at the beginning and the end of each permutation (and assumed to be fixed). This is due to the need 
to make the distinction between a permutation and its reverse order permutation. 

\bdefin \cite{IR}
The {\em $\M$-profile of a permutation} $P$ on $\n\cup\{0,n+1\}$ is the set of {\em $\M$-constraints}

$$\M(P)=\{t\frac{\hspace*{0.1cm}{\scriptstyle [\min _t, \max _t]}\hspace*{0.1cm}}{} t+1\, |\, 0\leq t\leq n\}$$

\noindent where $\min_t$ ($\max_t$ respectively) is the minimum (maximum respectively) element  
in the interval delimited on $P$ by the element $t$ (included) and the element $t+1$ (included). 

\edefin

Note that the relative positions on $P$ ({\em i.e.} which one is on the left of the other) of $t, t+1$ on the one hand, and of  $\min_t,\max_t$ on the other hand are not indicated by a $\M$-profile.
In the case where the relative positions of $t$ and $t+1$  are
known for all $t$, we use the term of {\em directed $\M$-profile} and the notations
$t\rf{[\min _t, \max _t]} t+1$ when $t$ is on the left of $t+1$,
respectively  $t\lf{[\min _t, \max _t]} t+1$ when $t+1$ is on the left of $t$.

\bex
Let $P=(0\, 6\, 4\, 7\, 2\, 9\, 1\, 8\, 5\, 3\, 10)$ a permutation on $[9]\cup\{0,10\}$. Then its
$\M$-profile is (note that the $\M$-constraints sharing an element are concatenated):

$$0\frac{\scriptstyle{[0, 9]}}{} 1\frac{\scriptstyle{[1, 9]}}{}2\frac{\scriptstyle{[1, 9]}}{}3\frac{\scriptstyle{[1, 9]}}{}4\frac{\scriptstyle{[1,9]}}{}5\frac{\scriptstyle{[1, 9]}}{}6\frac{\scriptstyle{[4, 7]}}{}7\frac{\scriptstyle{[1, 9]}}{}8\frac{\scriptstyle{[1, 9]}}{}9\frac{\scriptstyle{[1, 10]}}{}10$$

\noindent whereas its directed $\M$-profile is:

$$0\rf{[0, 9]} 1\lf{[1, 9]}2\rf{[1, 9]}3\lf{[1, 9]}4\rf{[1,9]}5\lf{[1, 9]}6\rf{[4, 7]}7\rf{[1, 9]}8\lf{[1, 9]}9\rf{[1, 10]}10$$

\noindent Notice that the $\M$-profile and the directed $\M$-profile of any permutation obtained by 
arbitrarily permuting the elements $\{3,5,8\}$ are the same, showing that a (directed or not) $\M$-profile
may correspond to several distinct permutations. 
\eex
The {\em $\M$-profile of a set $\PK$ of permutations} is defined similarly \cite{IR}, by requiring that $\min_t$ 
and $\max_t$ be defined over the union of the intervals delimited by $t$ (included) and $t+1$ (included)
on the $K$ permutations in $\PK$. This definition is given here for the sake of completeness,
but is little used in the paper.

We distinguish between the $\M$-profile of a (set of) permutation(s) and a {\em $\M$-profile}:

\bdefin
A {\em $\M$-profile} on $\n\cup\{0, n+1\}$ is a set of {\em $\M$-constraints} 

$$F=\{t\frac{\hspace*{0.1cm}\scriptstyle{[m_t, M_t]}\hspace*{0.1cm}}{}t+1\, |\, 0\leq t\leq n\}$$

\noindent with $0\leq m_t\leq t< t+1\leq  M_t\leq n+1$. 
\edefin

Again, a $\M$-profile is {\em directed} when for all $t$, $0\leq t\leq n$, the relative position of $t$ with respect to 
$t+1$ is given.
A $\M$-profile may be the $\M$-profile of some permutation, or of a set of permutations, but may  also be the profile 
of no (set of) permutation(s). We limit this study to one permutation, and therefore formulate the following problem:
\newpage

\noindent {\sc $\M$-Betweenness}

\noindent{\bf Input:} A positive integer $n$, a  $\M$-profile $F$ on $\n\cup \{0, n+1\}$. 

\noindent{\bf Question:} Is there a permutation $P$ on $\n\cup\{0,n+1\}$ whose $\M$-profile is $F$? 
\bigskip

The $\M$-{\sc Betweenness} problem is obviously related to the {\sc Betweenness} problem, since looking for a permutation
$P$ with $\M$-constraints defined by $F$ means satisfying a number of betweenness constraints.  Some differences
exist however, as $F$ also defines non-betweenness constraints. More precisely, each $\M$-constraint
$t\frac{\hspace*{0.1cm}\scriptstyle{[m_t, M_t]}\hspace*{0.1cm}}{}t+1$ from $F$ may be expressed using the betweenness constraints 
(abbreviated B-constraints):

\begin{equation}
\label{eq:Bc}
t\lrf{m_t}t+1, t\lrf{M_t}t+1
\end{equation}
\noindent along with the non-betweenness constraints (abbreviated NB-constraints):

\begin{equation}
\label{eq:NBc1}
\neg (t\lrf{j}t+1), j=0, 1, \ldots, m_t-1,M_t+1, \ldots, n+1.
\end{equation}

It is easy to imagine that in the $\M$-{\sc Betweenness} problem, the lack of information about
the relative position of  $t$ and $t+1$ on the permutation $P$ (i.e. which one is on the left of the other)
is a major difficulty. The directed version of the problem, in which these relative positions are given, should
possibly be easier.
\bigskip

\noindent{\sc Directed $\M$-Betweenness}

\noindent {\bf Input:} A positive integer $n$, a directed $\M$-profile $F$ on $\n\cup\{0, n+1\}$.

\noindent{\bf Question:} Is there a permutation $P$ on $\n\cup\{0,n+1\}$ whose directed $\M$-profile is $F$? 
\bigskip

\br
It is worth noticing here that in a (directed or not) $\M$-profile which corresponds to
at least one permutation on $\n\cup\{0,n+1\}$, the value $0$ ($n+1$ respectively) should
only occur in one precise $\M$-constraint, namely the one involving $0$ and $1$ ($n$ and
$n+1$ respectively). Otherwise, $0$ ($n+1$ respectively) cannot be the leftmost (rightmost, respectively)
value in the permutation. In the subsequent of the paper, it is assumed that this condition
has been verified before further investigations, and assume therefore that $0$ and $n+1$
are respectively located in places $0$ and $n+1$.
\label{rem:places0n+1}
\er

We present below, in Section \ref{sect:GeneralD}, our analysis of the Directed $\M$-{\sc Betweenness} problem,
proposing a first algorithmic approach and pointing out the main difficulties for reaching a complete polynomial solution. 
In Section \ref{sect:Particular}, we identify a polynomial particular case for the directed version. 
In Section \ref{sect:Generalizations} we propose to generalize $\M$-profiles to $\kM$-profiles, by introducing 
a parameter $k$ which allows to progressively increase the amount of information contained in a $\kM$-profile, 
up to a value $k_0$ which allows to identify each permutation by its $k_0$-profile. Section \ref{sect:Conclusion} is the
conclusion.

\section{Seeking an algorithm for {\sc Directed $\M$-Betweenness}}\label{sect:GeneralD}

\subsection{A na\"{\i}ve approach}

Let $F$ be a directed $\M$-profile on $\n\cup\{0,n+1\}$. The most intuitive idea for solving 
Directed $\M$-{\sc Betweenness} is  to build a simple
directed graph $G$ ({\em i.e.} with no loops or multiple arcs)
whose vertex set $V(G)$ is $\n\cup\{0,n+1\}$ and whose arcs $(x,y)$ indicate the precedence relationships
between the elements on each permutation corresponding to the given $\kM$-profile ({\em i.e.} $x$ is on the
left of $y$). If a permutation exists, $G$ must be a directed acyclic graph (or DAG). 
The $\M$-constraints from $F$ directly define arcs using: 1) the relative order between $t$ and $t+1$,  
for each $t\in \n$ (the corresponding arcs of $G$ are called {\em R-arcs}), 
and 2) the B-constraints (resulting into  {\em B-arcs}). Further arcs may be dynamically obtained by repeatedly invoking:  
3) the transitivity of the precedence relationship 
(resulting into {\em T-arcs}), and 4) the NB-constraints  (resulting into {\em NB-arcs}).  

Algorithm \ref{algo:Arcs} shows these steps. After the construction of the $R$- and $B$-arcs (steps
2-6), either transitivity or NB-constraints may be arbitrarily invoked to add supplementary
arcs as long as possible, performing what we call the {\em NB-transitive closure}
of $G$. This is done by the Build-Closure algorithm (Algorithm \ref{algo:Closure}), called in
step 8 of Algorithm \ref{algo:Arcs}. It is clear that in step 1 of the Build-Closure algorithm
a $T$-arc $(x,y)$ may be added iff there is a vertex
$c$  such that $(x,c)$ and $(c,y)$ are arcs, but $(x,y)$ is not an arc. The condition for
adding the $NB$-arc $(x,y)$ is slightly more complex, as $(x,y)$ may be added iff

\begin{algorithm}[t,boxed]
\caption{The Build-Easy-Arcs algorithm}
\begin{algorithmic}[1]
{\small \REQUIRE A directed $\M$-profile $F$ over $\n\cup\{0,n+1\}$.
\ENSURE  Either the answer "No" (meaning no permutation exists), or the pair 
$(G, \Sil)$ where $G$ is the DAG containing all deducible $R-$, $N-$, $T-$ and $NB$-arcs, and
$\Sil$ is the set of silent NB-constraints. \\
\hspace*{-0.6cm}({\sl  Note:} Arcs are added only if they do not create loops, nor multiple arcs with common source and target.)

\medskip

\STATE $G\leftarrow (\n\cup\{0,n+1\}, \emptyset)$
\FOR{each $t\in[n]\cup\{0\}$}
\STATE {\bf if} $t\rf{[m_t, M_t]} t+1\in F$ {\bf then} $tl\leftarrow t$; $tr\leftarrow t+1$ {\bf else}   $tl\leftarrow t+1$; $tr\leftarrow t$ {\bf end if}
\STATE add the R-arc $(tl,tr)$ to $G$
\STATE add the B-arcs $(tl,m_t), (m_t,tr),(tl,M_t), (M_t,tr)$ to $G$ {\sl \hfill // according to (\ref{eq:Bc})}
\ENDFOR
\STATE $\Sil\leftarrow$ the set of all NB-constraints $\neg(t\lrf{a}t+1)$ deduced from $F$  {\sl \hfill // according to (\ref{eq:NBc1})}
\STATE $G\leftarrow$ Build-Closure($G,\Sil$)
\STATE remove from $\Sil$ all NB-constraints $\neg(t\lrf{a}t+1)$ for which a setting is already found
\IF{$G$ is not a DAG}
\STATE output "No"
\ELSE
\STATE output ($G, \Sil$)
\ENDIF}
\end{algorithmic}
\label{algo:Arcs}
\end{algorithm}

\begin{algorithm}[t,boxed]
\caption{The Build-Closure algorithm}
\begin{algorithmic}[1]
{\small \REQUIRE A simple directed graph $G$ with vertex set $\n\cup\{0,n+1\}$, a set $NBc$ of NB-constraints on $\n\cup\{0,n+1\}$.
\ENSURE  The NB-transitive closure of $G$ using the NB-constraints in $NBc$. \\
\hspace*{-0.6cm}({\sl  Note:} Arcs are added only if they do not create loops, nor multiple arcs with common source and target.)

\medskip

\WHILE{a $T$-arc or an $NB$-arc $(x,y)$ may be added}
\STATE add $(x,y)$ to $G$ 
\ENDWHILE
\STATE output($G$)}
\end{algorithmic}
\label{algo:Closure}
\end{algorithm}

\begin{itemize}

\item[$\bullet$] either an NB-constraint $\neg(y\lrf{x}z)$ with $z\in\{y-1,y+1\}$ exists in $NBc$,  and $(x,z)$ is an arc,

\item[$\bullet$] or an NB-constraint  $\neg(x\lrf{y}z)$ with  $z\in\{x-1,x+1\}$ exists in $NBc$,  and $(z,y)$ is an arc.
\end{itemize}

\noindent Clearly, this na\"{\i}ve approach for {\sc $\M$-Betweenness} attempts to 
exploit all the $\M$-constraints. Unfortunately,  for some NB-constraints 
$\neg(t\lrf{a}t+1)$ Algorithm \ref{algo:Arcs} may provide no setting ({\em i.e.} neither the arcs $(t,a)$ 
and $(t+1,a)$, nor the arcs $(a,t)$ and $(a,t+1)$ are present in $G$), as shown below. 
These constraints are called {\em silent NB-constraints}, and are returned by the algorithm together with $G$, if $G$ is a DAG (step 13).

\bex
Let $F$ be defined on $[9]\cup\{0,10\}$ by the following $\M$-constraints:

$$0\rf{[0, 9]} 1\lf{[1, 9]}2\rf{[1, 9]}3\lf{[1, 9]}4\rf{[1,9]}5\lf{[1, 9]}6\rf{[4, 7]}7\rf{[1, 9]}8\lf{[1, 9]}9\rf{[1, 10]}10$$

\begin{figure}[t]
\vspace*{-1.5cm}
\begin{center}
\includegraphics[width=15cm]{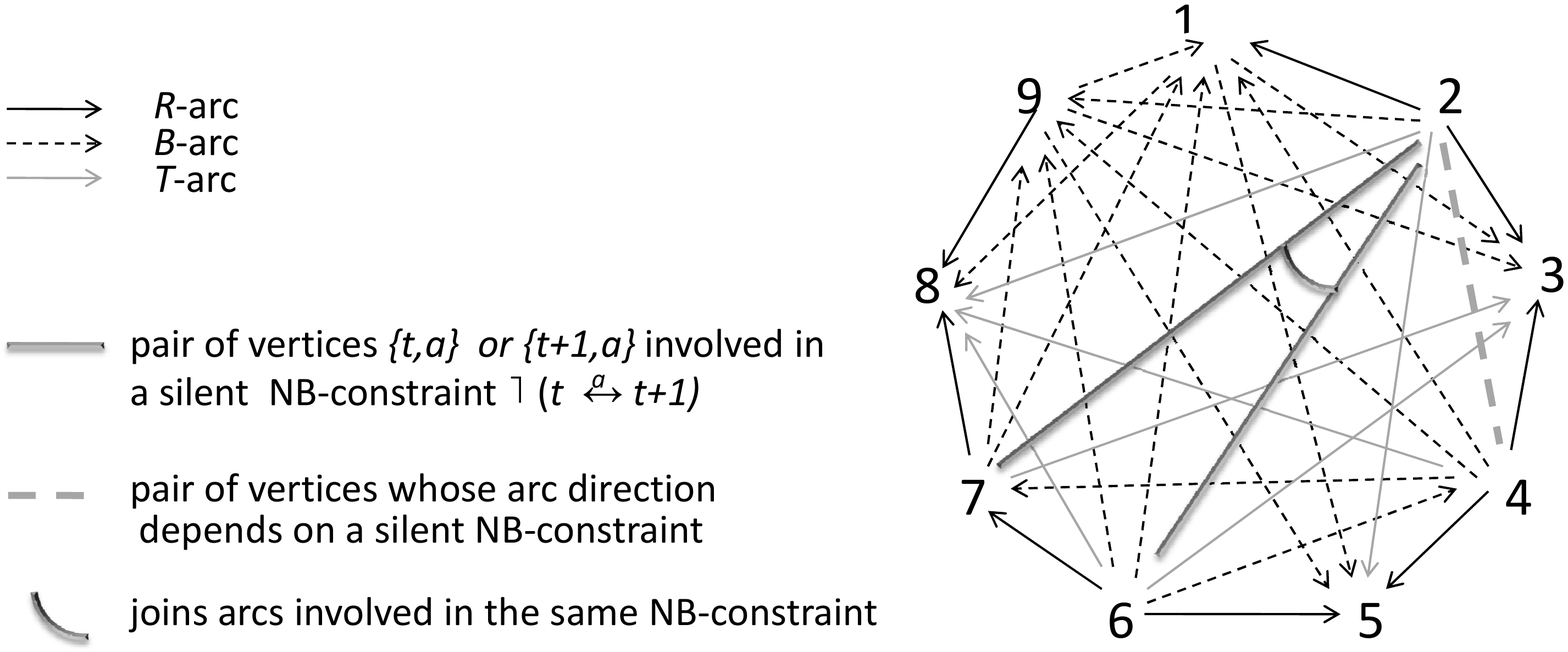}
\end{center}
\vspace*{-3cm}
\caption{{\small Directed acyclic graph $G$ obtained from Algorithm \ref{algo:Arcs} using the directed
$\M$-profile $0\rf{[0, 9]} 1\lf{[1, 9]}2\rf{[1, 9]}3\lf{[1, 9]}4\rf{[1,9]}5\lf{[1, 9]}6\rf{[4, 7]}7\rf{[1, 9]}8\lf{[1, 9]}9\rf{[1, 10]}10$. For simplicity reasons, vertices $0$ and $10$ and the arcs incident
to at least one of them are omitted. The pairs of vertices $\{2,4\}, \{2,6\}$ and $\{2,7\}$ are not arcs, they
show the silent NB-constraint  $\neg(6\xleftrightarrow{\,\,2\,\,} 7)$ and the pair $\{2,4\}$, that are related by 
coherent arc directions.  }}
\label{fig:ex}
\end{figure}

\noindent Figure \ref{fig:ex} shows the $R-$, $B-$ and $T-$ arcs used by Algorithm \ref{algo:Arcs} to
build the directed graph deduced from $F$. Vertices $0$ and $10$ are left apart in this figure, since the
constraints they are involved in allow only to place them at the beginning and respectively at
the end of the sought permutations.   The NB-constraints imposed by $F$ (except those with $0$ and $10$)
are $6\lrf{u}7$ with $u\in\{1, 2, 3, 8, 9\}$. When $u=1$ and $u=9$, both arcs involved in the NB-constraint are 
already in $G$ (due to $B$-constraints). For $u=3$ and $u=8$, all the arcs with $6$ and $7$ are built by
transitivity (although some of them may also be built using the appropriate NB-constraints), during the steps 8 
in Algorithm \ref{algo:Arcs}. For $u=2$, the NB-constraint cannot be used, since none of the arcs exists
(and no other arc may be created by transitivity). Then we
have $\Sil=\{\neg (6\lrf{2}7)\}$ at the end of Algorithm \ref{algo:Arcs}. 
Notice that the pairs $\{2,4\}, \{2,6\}$ and $\{2,7\}$ have correlated
directions in any setting, that is, either all three arcs have the source $2$,
or all three arcs have the target $2$. For $\{2,6\}$ and $\{2,7\}$ this is due
to the NB-constraint $\neg (6\lrf{2}7)$, whereas for $\{2,4\}$ this is due to the
transitivity ensured by the arcs $(6,4)$ and $(4,7)$.
\eex

Our problem is now this one:

\bigskip
\noindent {\bf (P)} Given $G$ and a set of silent NB-constraints, decide whether a setting is possible
for each silent NB-constraint such that the graph resulting by transitive closure is a DAG.
\bigskip

Unfortunately, the following result shows the difficulty of the problem:

\bfn \cite{guttmann2006variations}
Problem (P) is NP-complete even when the silent NB-constraints involve disjoint triples of vertices.
\label{claim:NPc}
\efn

Notice however that the graph $G$ we obtain at the end of Algorithm \ref{algo:Arcs} may have particular 
features (that we have not identified) making that we are dealing
with a particular case of problem (P). Claim \ref{claim:NPc} shows therefore that our problem is
potentially difficult, but does not prove its hardness.

\br
From an algorithmic point of view, we may notice that with the output of Algorithm \ref{algo:Arcs}
we may easily find a parameterized algorithm for $\M$-{\sc Betweenness}. Given $G$ and $\Sil$,
we have $O(2^{|\Sil|})$ possible settings to test, thus resulting into 
an FPT algorithm with parameter $s$ given by the number of silent NB-constraints. 
\er

\subsection{Further analysis of arc propagation}

With the aim of forcing the setting of some appropriately chosen silent NB-constraint,  
let us now analyze the impact of adding an arbitrary arc $(a_1,b_1)$ to $G$, where  
$a_1$ and $b_1$  are non-adjacent vertices from $G$. Denote $G+(a_1,b_1)$  the graph 
obtained from $G$ be adding the arc $(a_1,b_1)$, and  let $G^1$ be the NB-transitive closure of 
$(G+(a_1,b_1), \Sil)$,  {\em i.e.} the directed graph obtained by performing 
Build-Closure$(G+(a_1,b_1), \Sil)$. 

Several definitions are needed before going further. Given an NB-constraint  $\neg(t\lrf{a}t+1)$, the vertex
$a$ of $G$ is called the {\em top} of the NB-constraint, whereas the pair $\{t,t+1\}$  is called
the {\em basis} of the NB-constraint. An arc $(x,y)$ is {\em new} if it is an arc of $G^1$ but
not of $G$, and is {\em old} if it is an arc of $G$. New arcs are obtained using 
Build-Closure$(G+(a_1,b_1), \Sil)$ according to a certain linear order, resulting from the arbitrary choices made in step 1. This order is denoted $\alpha$, such that $(x_1,y_1)\alpha
(x_2,y_2)$ means that $(x_1,y_1)$ is created by Build-Closure before $(x_2,y_2)$.
Then, the following claim is simple:

\bfn
For each new arc $(v,w)$, there exists a series of new arcs $U: =(v_1,w_1), (v_2,w_2),$ $ \ldots, (v_z,w_z)$ such that  $(v_1,w_1)=(a_1,b_1)$,  $(v_z,w_z)=(v,w)$, $(v_i,w_i)\alpha(v_{i+1},w_{i+1})$ for all $i$ with $1\leq i\leq z-1$ 
and each arc $(v_{i+1}, w_{i+1})$, $1\leq i\leq z-1$, is obtained from the preceding one $(v_i,w_i)$ using one of the following cases:

\begin{enumerate}
\item $w_{i+1}=w_i$ and $(v_{i+1},v_i)$ is either an old arc, or a new arc such that $(v_{i+1},v_i)\alpha (v_{i+1}, w_{i+1})$; in this case $(v_{i+1},w_{i+1})$ is a new $T$-arc.
\item  $w_{i}=w_{i+1}$ and $\{v_i,v_{i+1}\}$ is the basis of an NB-constraint of $\Sil$ 
with top $w_i$; in this case $(v_{i+1},w_{i+1})$ is a new $NB$-arc.
\item $v_i=v_{i+1}$ and $(w_i,w_{i+1})$ is either an old arc,  or a new arc such that
$(w_{i},w_{i+1})\alpha (v_{i+1}, w_{i+1})$; in  this case $(v_{i+1},w_{i+1})$ is a new $T$-arc.
\item  $v_i=v_{i+1}$ 
and $\{w_i,w_{i+1}\}$ is the basis of an NB-constraint from $\Sil$ with top $v_i$;
in this case $(v_{i+1},w_{i+1})$ is a new $NB$-arc.
\end{enumerate}
\label{claim:imminduce}

\efn

{\bf Proof.} In order to obtain  $(v_{i+1},w_{i+1})$, we need to apply either the transitivity
(step 2 in Algorithm \ref{algo:Closure} for a $T$-arc,  which gives cases 1 and 3), or an NB-constraint from $\Sil$
(again step 2 in Algorithm \ref{algo:Closure}, but for an $NB$-arc, which gives cases 2 and 4). 
$\Box$  
\bigskip

The sequence $U$ is called a {\em setting sequence} for $(v,w)$, whereas the index $i$
of an arc $(v_i,w_i)$ is called its {\em range} in $U$. From now on, the
case in Claim \ref{claim:imminduce} used to deduce one arc from the preceding one in a setting sequence 
is  indicated between the two arcs.

\bex
For the example in Figure \ref{fig:ex}, if $a_1=2$ and $b_1=7$, then $U: =(2,7) \frac{4}{} (2,6)\frac{3}{}(2,4)$
is a setting sequence for $(2,4)$ using case 4 followed by case 3 in Claim \ref{claim:imminduce} to go
from one arc to the next one. 
\label{ex:U}
\eex

Now, let $a_1, a_2, \ldots, a_s$ (respectively  
$b_1, b_2, $ $\ldots,$ $b_r$) be the subsequence of $v_1, \ldots, v_z$ 
(respectively of $w_1, \ldots, w_z$) obtained by replacing  {\em consecutive} 
copies of the same vertex with only one copy of that vertex. Equivalently, if $(a_i,b_j)$ is 
an arc of $U$, then the next arc is either  $(a_{i+1}, b_j)$ (cases 1 and 2 in 
Claim \ref{claim:imminduce}) or $(a_i,b_{j+1})$ (cases 3 and 4 in Claim
\ref{claim:imminduce}). Of course, we have $a_1=v_1, a_s=v_z=v, b_1=w_1$ and $b_r=w_z=w$.

\begin{figure}[t]
\vspace*{-1.5cm}
\begin{center}
\includegraphics[width=15cm]{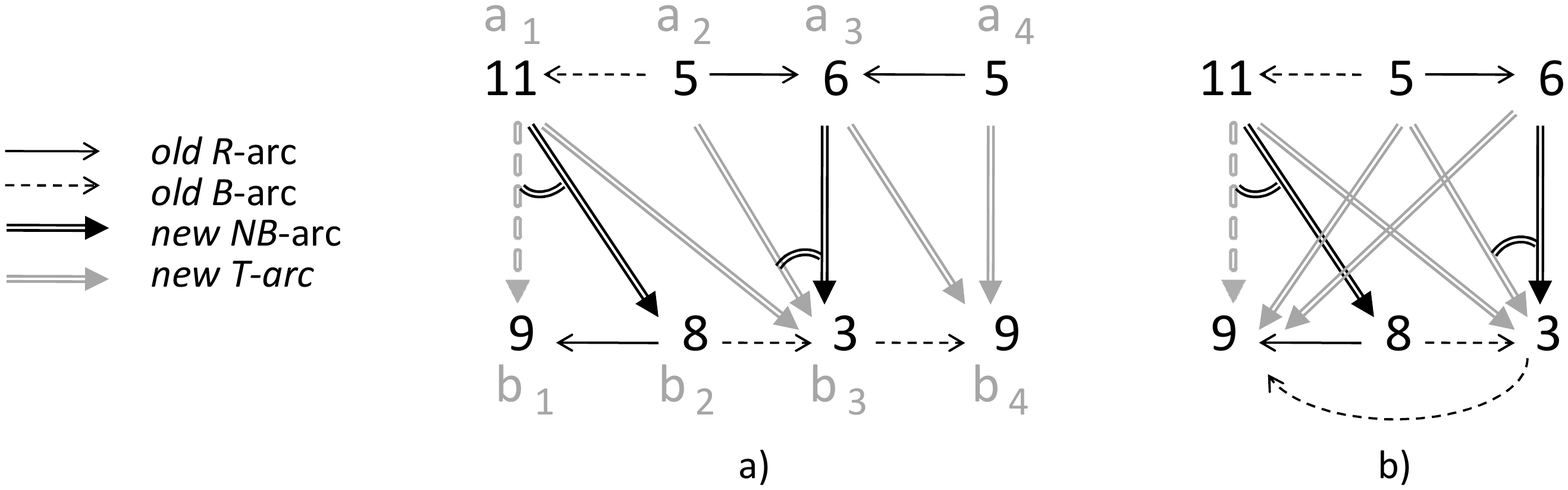}
\end{center}
\vspace*{-4cm}
\caption{{\small Old arcs (thin horizontal arrows) and new arcs (thick vertical arrows) used by 
the setting sequence $U: =(11,9)\frac{4}{} (11,8)\frac{3}{} (11,3)\frac{1}{} (5,3)\frac{2}{} (6,3)\frac{3}{} (6,9)\frac{1}{} (5,9)$ in Example \ref{ex:HU}. a) Using notation $a_i$, $1\leq i\leq 4$, and $b_j$, $1\leq j\leq 4$. b) Using 
the vertices in $G$, and thus defining the graph $H_{U}$. In both cases, the initial arc $(a_1,b_1)$ is
a dotted arrow.}}
\label{fig:ex3}
\end{figure}

\bex
Consider $P=(0\,\, 7\,\, 4\,\, 10\,\, 2\,\, 1\,\, 12\,\, 8\,\, 3\,\, 9\,\, 5\,\, 11\,\, 6\,\, 13)\}$,
and let $F$ be the $\M$-profile of $P$. For $F$, apply Algorithm \ref{algo:Arcs} to obtain the
graph $G$ and the set $\Sil$.  Then $G$ - that the reader is invited to build it himself - is partitioned into three sets, respectively made of:
the vertices preceding the pair $\{1,12\}$, the pair $\{1,12\}$ (in this order, and with no intermediate vertex),
and the  vertices following the pair $\{1,12\}$. The set $\Sil$ is $\{8\lrf{11}9, 5\lrf{3}6\}$, and thus
involves only vertices in the third set, which induces in
$G$ the subgraph $G'$ with vertex set $\{3, 5, 6, 8, 9, 11\}$ and arcs
$\{(8,3), (3,9), (8,9)\}\cup\{(5,11), (11,6), (5,6)\}$.
With $(a_1,b_1)=(11,9)$, 
we have (see Figure \ref{fig:ex3}a) that $U: =(11,9)\frac{4}{} (11,8)\frac{3}{} (11,3)\frac{1}{} (5,3)\frac{2}{} (6,3)\frac{3}{} (6,9)\frac{1}{} (5,9)$ is a setting sequence
for $(5,9)$ with $a_1=11, a_2=5, a_3=6, a_4=5$ (thus $s=4$) and $b_1=9, b_2=8, b_3=3, b_4=9$ (thus $r=4$).
\label{ex:HU}
\eex

\br
Notice that we could possibly have $a_i=a_l$, for distinct $i,l\in\{1, 2, \ldots, s\}$, {\em i.e.} they
correspond to the same vertex of $G$), if two
arcs with the same endpoint are set in distant steps of the setting process represented by $U$. We could also
possibly have $a_i=b_j$  for some $i,j$ if, for instance, $a_1, \ldots, a_h$ (with $h> i$) are distinct,
$b_1, \ldots, b_{j-1}$ are  distinct, $(a_h,b_{j-1})$ is a new arc and $(b_{j-1},a_i)$ is an old arc 
(making that the vertex $b_{j}$ is equal to $a_i$, and thus by transitivity - or case 3 in Claim \ref{claim:imminduce} - 
one sets $(a_h,a_i)$).
\label{rem:several}
\er

\br
Also note that for every pair of  arcs $(a_i,b_j)$ and $(a_{p},b_{q})$ from $U$, we
have either $i\leq p$ and $j\leq q$ (when $(a_i,b_j)\alpha(a_p,b_q)$), or $i\geq p$ 
and $j\geq q$ (when $(a_p,b_q)\alpha(a_i,b_j)$). It is therefore understood,
here and in the subsequent of the paper, that in case $a_i=a_l$ for some $i\neq l$,
we make a clear difference between the arcs $(a_i,b_j)$ of $U$ and the arcs $(a_l, b_f)$ of $U$.
These arcs are incident with the same vertex of $G$ but this vertex is called $a_i$
in the first case, and $a_l$ in the second one.
\label{rem:HU}
\er

\noindent {\bf Example \ref{ex:HU} (cont'd)}. We have
$a_2=a_4=5$ and $b_1=b_4=9$, but when we refer to the new arcs containing $a_2$ we only refer to
the arc $(5,3)$ and when we refer to the new arcs containing $a_4$ we only refer to the arc $(5,9)$.
Similarly, when we refer to the new arcs incident with $b_1$ we refer only to the arc $(11,9)$ whereas
when we refer to those incident with $b_4$ we mean the arcs $(6,9)$ and $(5,9)$.
\bigskip

In order to represent arc propagation, we need to look closer to the partial subgraph  $H_U$ of $G^1$ 
given by the set of {\em distinct} vertices  used in the
setting sequence $U$, the arcs in $U$ and the arcs used to deduce each arc of $U$ from the previous one, 
using Claim \ref{claim:imminduce}. The graph $H_U$ is defined as:
\medskip

$V(H_U)=\{a_i\in V(G)\, |\, 1\leq i\leq s\}\cup\{b_j\in V(G)\, |\, 1\leq j\leq r\}$ 

$E(H_U)=U\cup \{(a_{i+1}, a_i) \,|\ \exists b_j : (a_{i+1},b_j)\, \mbox{is deduced from}\, (a_i,b_j)\, \mbox{in}\ U\, \mbox{using case 1}\}$

\hspace*{1.85cm} $\cup \{(a_{i}, a_{i+1})\,|\ \exists b_j : (a_{i+1},b_j)\, \mbox{is deduced from}\, (a_i,b_j)\, \mbox{in}\ U\, \mbox{using case 2}\}$

\hspace*{1.9cm}$\cup \{(b_{j}, b_{j+1})\,|\ \exists a_i : (a_{i},b_{j+1})\, \mbox{is deduced from}\, (a_i,b_j)\, \mbox{in}\ U\, \mbox{using case 3}\}$

\hspace*{1.9cm}$\cup \{(b_{j+1}, b_{j})\,|\ \exists a_i : (a_{i},b_{j+1})\, \mbox{is deduced from}\, (a_i,b_j)\, \mbox{in}\ U\, \mbox{using case 4}\}$

\medskip

\noindent The graph $H_U$ is called the {\em setting path} associated with $U$ (or, alternatively, a {\em setting path}
for $(a_s,b_r)$). Notice that case 2 (respectively case 4) in Claim \ref{claim:imminduce} may be included in case 1 
(respectively case 3) when the basis is the arc $(a_{i+1},a_i)$ (arc $(b_j, b_{j+1})$ respectively). The definition of
$H_U$ keeps as case 2 (respectively case 4) only the configuration not included in case 1 (respectively case 3).
See Figure \ref{fig:ex3}b).

Claim \ref{claim:imminduce} and the definition of $H_U$ allow us to have a basis for 
future analysis, but also show us that the
choice of one arc $(a_1,b_1)$ has effects that are difficult to measure accurately. 
The NP-completeness of the problem {\bf (P)} (see Claim \ref{claim:NPc}) comes from 
this complex arc propagation, which makes that different setting sequences with the
same initial arc may lead to conflicts, {\em i.e.} to circuits.

\begin{figure}[t]
\vspace*{-1.5cm}
\begin{center}
\includegraphics[width=15cm]{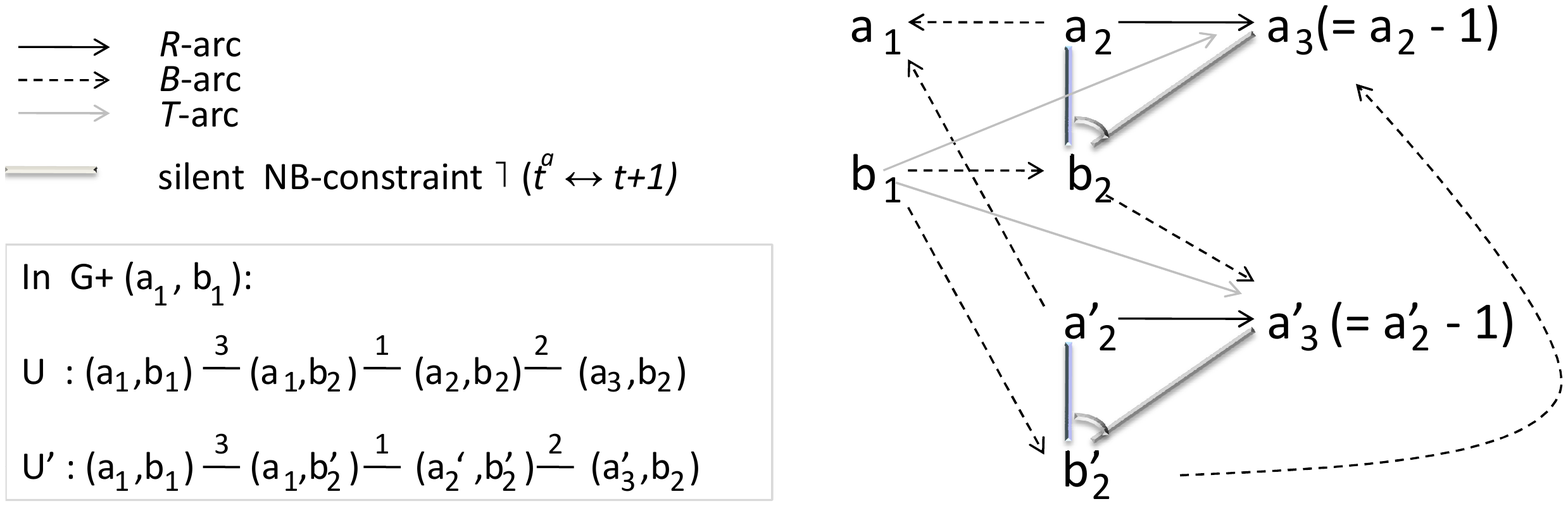}
\end{center}
\vspace*{-4cm}
\caption{{\small  Configuration where the addition of  the arc $(a_1,b_1)$ sets
(among other arcs) the arcs $(a_3,b_2)$ and $(a'_3, b'_2)$, thus inducing a circuit.
The types of the arcs are the same as in Figure \ref{fig:ex}, but the pairs depending
on the setting of a silent NB-constraint are missing.}}
\label{fig:ex2}
\end{figure}

\bex
In Figure \ref{fig:ex2}, we present a configuration (which is a subgraph of $G$) showing that not each possible
setting is a correct setting, since imposing the existence of one arc $(a_1,b_1)$ may induce
circuits in the graph $G^1$. In this configuration, setting the arc $(a_1,b_1)$ implies the additional
arcs $(a_3,b_2)$ and $(a'_3,b'_2)$, and thus the construction of a circuit. A $\M$-profile inducing such a configuration in the
associated DAG $G$ is the following one (where $a_1=18, b_1=12, a_2=22, a_3=21,
a'_2=16, a'_3=15, b_2=25, b'_2=8$):
\medskip

$0\rf{[0,27]}1\rf{[1,29]}2\lf{[1,29]}3\rf{[1,29]}4\lf{[1,29]}5\rf{[1,29]}6\lf{[1,29]}{\bf 7\lf{[3,21]}8}
\rf{[1,29]}9\lf{[1,29]}10$  

$10\rf{[1,29]}11\lf{[1,29]}{\bf 12\rf{[8,25]}13}\rf{[1,29]}14\lf{[1,29]}{\bf 15\lf{[10,27]}16}\rf{[1,29]}17\lf{[1,29]}18$

${\bf 18\lf{[16,22]}19} \rf{[1,29]}20\lf{[1,29]}{\bf 21\lf{[5,22]}22}\rf{[1,29]}23\lf{[1,29]}{\bf 24\lf{[15,25]}25}\rf{[1,29]}26$

$\lf{[1,29]}27\rf{[1,29]}28\lf{[2,29]}29\rf{[2,30]}30$
\medskip

\noindent In this example, the $\M$-constraints in bold define the arcs needed to obtain
the configuration in Figure \ref{fig:ex2}, and some additional arcs. The elements
involved in these $\M$-constraints are, in every permutation with this $\M$-profile,
on the left of $1$ and $29$ (the minimum and maximum elements), which are neighbors and
in this order on each permutation. The remaining of the elements are on the right of $1$ 
and $29$, and are intended to complete the set $\{1, 2, \ldots, 30\}$ without any
participation to the configuration.
\eex

In order to find polynomial particular cases, we need to be able to control the
form of the setting paths, and this is what we do in the subsequent. To this end,
notice that:

\br
The vertices $0$ and $n+1$ belong to no setting path. Indeed, according to Remark \ref{rem:places0n+1}, it is assumed that they are definitely located at places
$0$ and $n+1$ respectively, and thus their relative positions with respect to any other
element are known. No arc incident to any of them may thus be added, as would be the
case if they belonged to some setting path.
\label{rem:no0n+1} 
\er

\section{Polynomial case for {\sc Directed $\M$-Betweenness}}\label{sect:Particular}

Say that a $\M$-profile $F$ on $\n\cup\{0,n+1\}$ is {\it linear} if the inclusion between sets defines a linear order on the intervals
$[m_t..M_t]$, $1\leq t\leq n-1$, where the notation $(a..b)$ denotes the set of integers $x$ with $a\leq x\leq b$.
We show in this section that the problem {\sc Directed $\M$-Betweenness} is polynomial for linear $\M$-profiles.

Given $c\in [n]$, let $NB(c)=\{t\, |\, 1\leq t\leq n-1, c< m_t\, \mbox{or}\, M_t<c\}$. In other words, $NB(c)$ is the set of values $t$ such that
$\{t,t+1\}$ is the basis of an NB-constraint with top $c$.

\bfn
Let $F$ be a linear profile on  $[n]\cup\{0,n+1\}$. Then the inclusion between sets defines a linear order denoted $\prec$ on
the sets $NB(c)$, $1\leq c\leq n$.
\label{claim:totalNB}
\efn

{\bf Proof.} By contradiction, assume that $c_1$ and $c_2$ exist such that $NB(c_1)\setminus NB(c_2)$ contains
$t_1$ and $NB(c_2)\setminus NB(c_1)$ contains $t_2$. Then $t_1\neq t_2$. 

In the case where $c_1<m_{t_1}$ and $c_2<m_{t_2}$, assume w.l.o.g. that $c_1<c_2$. Then $c_1<m_{t_2}$
and thus $t_2\in NB(c_1)$, a contradiction.  The case where  $c_1>M_{t_1}$ and $c_2>M_{t_2}$
is similar.

In the case where $c_1<m_{t_1}$ and $c_2>M_{t_2}$,
recall that by hypothesis $F$ is linear, and thus either $[m_{t_1}..M_{t_1}]\subseteq [m_{t_2}..M_{t_2}]$ or vice-versa.
If $[m_{t_1}..M_{t_1}]\subseteq [m_{t_2}..M_{t_2}]$, then $m_{t_2}\leq m_{t_1}< M_{t_1}\leq M_{t_2}$
and with $c_2>M_{t_2}$ we deduce that $t_1\in NB(c_2)$, a contradiction. If $[m_{t_2}..M_{t_2}]\subseteq [m_{t_1}..M_{t_1}]$, then $m_{t_1}\leq m_{t_2}< M_{t_2}\leq M_{t_1}$
and with $c_1<m_{t_1}$ we deduce that $t_2\in NB(c_1)$, a contradiction. $\Box$

\bigskip

Now, assume Algorithm \ref{algo:Arcs} has been applied for $F$, and let $(G, \Sil)$ be its output, assuming $G$ is a DAG.
To finish the algorithm for $F$, we apply Algorithm \ref{algo:linear}. The following claim is easy but very useful.

\bfn
The vertex $b_1$ chosen in Algorithm \ref{algo:linear} has the following properties:

\begin{itemize}
\item[$(a)$] $NB(b_1)$ is maximum with respect to the linear order $\prec$ on the set

$\{NB(c)\, |\, 1\leq c\leq n\, \mbox{and}\, c\, \mbox{is the top of at least one constraint from}\, \Sil\}$.
\item[$(b)$] $b_1$ does not belong to a basis, but is a top for all the basis defining constraints from $\Sil$.
\end{itemize}

\label{claim:b1}
\efn

{\bf Proof.} The first affirmation is clear by the choice of $b_1$ in step 3 of the algorithm and 
Claim \ref{claim:totalNB}. The second affirmation is deduced by contradiction. If $b_1$
belonged to a basis $\{b_1,b_1+1\}$ or $\{b_1-1, b_1\}$ with top $c$, then we would have $NB(c)\not\subseteq NB(b_1)$
since the basis $\{b_1,b_1+1\}$ or $\{b_1-1, b_1\}$ cannot have top $b_1$ (the vertices of a basis are by definition distinct from
its top). The second part of affirmation $(b)$ results directly from affirmation $(a)$. $\Box$
\bigskip

In the next claims, we show the correctness of our algorithm.
To this end, each arc $(a_i,b_j)$ of $U$ (and thus of $H_U$) is called a {\em local new arc} 
with respect to $U$,  in order to make the difference with the arcs from $G^1$ which are new but 
do not belong to $U$, termed {\em non-local new arcs}. Similarly, a vertex $a_i$ of $H_U$ is 
a {\em local top} if there  exists $b_q$ such that  
$\neg(b_q\lrf{a_i}b_{q+1})$ is an NB-constraint used by $U$, i.e. one of the arcs $(a_i,b_q)$ and $(a_i, b_{q+1})$
is deduced from the other in $H_U$, using case 4.  The pair $\{b_q,b_{q+1}\}$ is in this case a {\em local basis}. 
The symmetric definitions hold for a vertex $b_j$ (instead of $a_i$). Note that a local basis has a unique local top,
by Remark \ref{rem:HU}.

For any vertex  $a_i$, we also denote $first_U(i)$ the minimum $u$ with $1\leq u\leq r$
such that $(a_i,b_u$ belongs to $U$.

\begin{algorithm}[t,boxed]
\caption{The Linear-Profile algorithm}
\begin{algorithmic}[1]
\REQUIRE The output $(G, \Sil)$ of Algorithm \ref{algo:Arcs} for a directed linear $\M$-profile $F$ on $[n]\cup\{0,n+1\}$. 
\ENSURE  A permutation $P$ with the $\M$-profile $F$.

\WHILE{$\Sil\neq\emptyset$}
\STATE $C\leftarrow\{c\in\n\, |\, c\, \mbox{is the top of at least one constraint from}\, \Sil\}$
\STATE Choose $b_1\in C$ s.t. $|NB(b_1)|=max\{|NB(c)|\, |\, c\in C\}$
\STATE Choose $a_1$ such that $\neg(a_1\lrf{b_1}a_1+1)\in \Sil$. 
\STATE $G\leftarrow$ Build-Closure$(G+(a_1,b_1),\Sil)$
\ENDWHILE
\STATE $P\leftarrow$ topologically sort $G$
\STATE Output($P$)
\end{algorithmic}
\label{algo:linear}
\end{algorithm}

\bfn
Let $F$ be a directed linear profile on  $[n]\cup\{0,n+1\}$ and
let  $a_1$ and $b_1$ be chosen as in  Algorithm \ref{algo:linear}. Then the following
affirmations hold:

\begin{itemize}

\item[$(a)$] Let $(v,w)$ be a new arc of $G^1$ and let $U$ be setting sequence 
for $(v,w)$ with arc sources  $a_1, a_2, \ldots, a_s$ and arc targets $b_1, b_2, \ldots, b_r$.
Then there is no old arc $(b_1,a_i)$ in $G^1$, with $1< i\leq s$.
\item[$(b)$] All arcs $(b_1,x)$ of $G^1$ are old.
\end{itemize}
\label{claim:before5}
\efn

{\bf Proof.} To prove (a), we assume by contradiction that the affirmation is false,
and choose $(v,w), U$ and $(b_1,a_i)$ such that the arc $(a_i,b_{first_U(i)})$ is the smallest 
with respect to the order $\alpha$. Several cases occur.

\begin{itemize}
\item[{\sl i)}] If $\{a_{i-1},a_i\}$ is a local basis (case 2 in the definition of $H_U$), 
then $b_1$ is also a top of it (by Claim \ref{claim:b1}(b)
and thus from the old arc $(b_1,a_{i})$ we deduce the existence of the old arc $(b_1,a_{i-1})$
(computed by the call of Build-Closure in step 8 of Algorithm \ref{algo:Arcs}). But then
the choice of $(b_1,a_i)$ is contradicted, since $(a_{i-1},b_{first_U(i-1)})\alpha (a_i,b_{first_U(i)})$.

\item[{\sl ii)}] If $(a_i,a_{i-1})$ is an old arc (case 1 in the definition of $H_U$, with an old arc), 
then $(b_1,a_{i-1})$ is also an old arc, computed by the call of Build-Closure in step 8 of Algorithm \ref{algo:Arcs}. As before, the choice of $(b_1,a_i)$ is contradicted.

\item[{\sl iii)}] If $(a_i,a_{i-1})$ is a local new arc (case 1 in the definition of $H_U$, with a local new 
arc), then this arc belongs to $U$ and was built before $(a_i, b_{first_U(i)})$ since
it must be built before its use. Then there exist $p,q$ with $1\leq p< i-1$ and 
$1\leq q\leq first_U(i)$ such that $(a_p,b_q)$ and $(a_i, a_{i-1})$ are the same arc,
but with different notations due to its multiple use in $H_U$ (see Remark  
\ref{rem:several}). In particular, $a_p$ and $a_i$ are the same vertex of $H_U$, and
thus $(b_1,a_p)$ is an old arc of $H_U$, with $p<i$. Once again, the choice of
$(b_1,a_i)$ is contradicted, since $(a_{p},b_{first_U(p)})\alpha (a_i,b_{first_U(i)})$.

\item[{\sl iv)}]  Finally, if $(a_i,a_{i-1})$ is a non-local new  arc (case 1 in the definition of $H_U$, with a non-local new arc), 
then it was built before $(a_i, b_{first_U(i)})$. Consequently, 
there exists a setting sequence $T$ for $(a_i,a_{i-1})$
with arc sources $c_1(=a_1), c_2, \ldots, c_g=a_i$ and arc targets 
$d_1(=b_1), d_2, \ldots, d_h=a_{i-1}$. In this setting sequence, we have that 
$(b_1,c_g)$ is an old arc, and $(c_g,d_h)=(a_i,a_{i-1})$. Then, 
$(c_g, d_{first_T(g)})\alpha (a_i,a_{i-1})\alpha$ $\alpha(a_i, b_{first_U(i)})$,
contradicting again the choice of $U$ and $(b_1,a_i)$.
\end{itemize}

To prove $(b)$, assume by contradiction that some arcs $(b_1,x)$ are  created by 
Build-Closure$(G+\{a_1,b_1\},\Sil)$, and let $(b_1,x_1)$ be the smallest of them according
to the order $\alpha$. Then in a setting sequence $U$ for $(b_1,x_1)$ with arc sources  
$a_1, a_2, \ldots, a_s$ and arc targets $b_1, b_2, \ldots, b_r$, we have $(b_1,x_1)=(a_p,b_q)$
for some $p,q$ with $1< p\leq s$ and $1< q\leq r$. Then the pair $\{a_{p-1},a_p\}$ is
not a basis since $a_p=b_1$ and by Claim \ref{claim:b1}(b), $b_1$ belongs to no basis. 
Then, $(a_p,a_{p-1})$ is an arc. This arc cannot be old, since then recalling that
$a_p=b_1$ we have that $(b_1,a_{p-1})$ is on old arc thus contradicting affirmation (a).
Then $(a_p,a_{p-1})$ must be a new arc. Now, we have by case 1 in Claim \ref{claim:imminduce}  that $(a_p,a_{p-1})\alpha (a_p,b_{first_U(p)})\alpha (a_p,b_q)$. Since $(a_p,a_{p-1})=(b_1,a_{p-1})$ and $(a_p,b_q)=(b_1,x_1)$
we deduce that $(b_1,a_{p-1})\alpha (b_1,x_1)$, thus contradicting the choice of $(b_1,x_1)$. $\Box$
\bigskip
Say that a setting sequence $U$ for $(v,w)$ with arc sources $a_1, a_2, \ldots, a_s$ and arc targets $b_1, b_2, \ldots, b_r$ is {\em canonical} if $H_U$ has the following properties:

\begin{enumerate}
\item[$(a)$] $b_1$ and (if it exists) $b_2$ are distinct from $a_i$, $1\leq i\leq s$, and $(b_1,b_2)$ is an old arc.
\item[$(b)$] $r\leq 2$.  
\item[$(c)$] $(a_i,b_1)\in U$, for all $i$ with $1\leq i\leq s$. 
\end{enumerate} 

\bfn
Let $F$ be a directed linear profile on  $[n]\cup\{0,n+1\}$ and
let  $a_1$ and $b_1$ be chosen as in  Algorithm \ref{algo:linear}. 
Let $(v,w)$ be a new arc of $G^1$.  Then, for each setting sequence $U$ for $(v,w)$ with
arc sources $a_1, a_2, \ldots, a_s$ and arc targets $b_1, b_2, \ldots, b_r$,
there is a canonical setting sequence $U^0$ for $(v,w)$ with arc sources 
$a_1, a_2, \ldots, a_s$ and arc targets $b_1$, and (whenever $b_1\neq b_r$) 
$b'_2=b_r$. 
\label{claim:shortpath}
\efn

{\bf Proof.} The proof is by induction on the range $k$ of $(v,w)$ (or, equivalently, of
$(a_s,b_r)$) in a setting sequence $U$ for $(v,w)$. Recall that the arc with range 1 
is $(a_1,b_1)$.

In the case $k=2$, we have either $r=1$ (when cases 1 or 2 in Claim 
\ref{claim:imminduce} are used to obtain the second arc), or $s=1$ (when cases 3 or 4 
are used).  When $r=1$ we are already done. When $s=1$, by Claim \ref{claim:before5}(b)
we know that $(b_1,b_2)$ is an old arc, and we are done.

In the general case, assume by inductive hypothesis that the claim holds for all
arcs with range less than $k$ in some setting sequence, and that the range of  $(v,w)$
(or, equivalently, of $(a_s,b_r)$) in $U$ is $k$.  We have two cases.

Case A. The arc preceding $(a_s,b_r)$ in $U$ is $(a_s,b_{r-1})$. By inductive hypothesis, for
$(a_s,b_{r-1})$ there is a canonical setting sequence $U^0:=(a_1,b_1), (a_2,b_1), \ldots ,(a_s,b_1)$ and 
(if $b_{r-1}\neq b_1$) $(a_s,b'_2)$, meaning that $b'_2=b_{r-1}$ when $b'_2$
exists, and $b_1=b_{r-1}$ when $b'_2$ does not exist. We have two (sub)cases:

\begin{enumerate} 
\item[A.1.]  When $b'_2$ exists, we have that $U^1:=U^0.(a_s,b_r)$ 
(this is concatenation)
is a setting sequence for $(a_s,b_r)$, in which  $(a_s,b_r)$ is obtained from $(a_s, b'_2)$ 
using the
same case of Claim \ref{claim:imminduce} as used in $U$. Notice that the case 3 with a new
arc $(b'_2, b_r)$ cannot appear, since then in any setting sequence $U'$ for 
$(b'_2, b_r)$ with arc sources $c_x$ and arc targets $d_y$, we have that 
$(b'_2, b_r)=(c_i, d_j)$ for some $i$ and $j$, implying that $(b_1, c_i)$ is an
old arc (as $c_i=b'_2$), a contradiction with Claim~\ref{claim:before5}(a). Then
only case 3 with an old arc, and case 4 may occur. Both cases imply that $(b_1,b_r)$
is an old arc, as follows. In case 3 with an old arc $(b'_2,b_r)$, the transitivity
using the old arc $(b_1, b'_2)$ implies indeed the construction of $(b_1,b_r)$
in step 8 of Algorithm \ref{algo:Arcs}. If $\{b'_{2},b_r\}$ is a
local basis ({\em i.e.} case 4 is used), we deduce that $b_1$ is a top for it, by
Claim \ref{claim:b1}(b). Now, since $(b_1,b'_2)$ is an old arc by inductive hypothesis,
we deduce that $(b_1,b_r)$ is also an old arc obtained from the NB-constraint with top
$b_1$ and basis  $\{b'_{2},b_r\}$. Thus $(b_1,b_r)$ is an old arc in all cases. Then
$U^2=(a_1,b_1), \ldots, (a_s,b_1), (a_s,b_r)$ is a setting sequence for $(a_s,b_r)$,
which is canonical if we ensure that $b_r$ is distinct from all $a_i$, $1\leq i\leq s$.
This is guaranteed by Claim \ref{claim:before5}(a).
\item[A.2.] When $b'_2$ does not exist, we have that $U^1: =U^0.(a_s,b_r)$ is
a canonical setting sequence for $(a_s,b_r)$. Indeed, as $b_1=b_{r-1}$ we know that
$(a_s,b_r)$ is obtained from $(a_s, b_1)$ using case 3 or 4 in Claim 
\ref{claim:imminduce}. Moreover, by Claim \ref{claim:b1}(b), $b_1$ belongs to no basis, 
thus $(b_1,b_2)$ is an old or new arc. But the latter possibility is forbidden by 
Claim \ref{claim:before5}(b).
\end{enumerate}

Case B. The arc preceding $(a_s,b_r)$ in $U$ is $(a_{s-1},b_r)$. By inductive hypothesis, for
$(a_{s-1},b_r)$ there is a canonical setting sequence $U^0:=(a_1,b_1), (a_2,b_1), \ldots ,(a_{s-1},b_1)$ and (if $b_r\neq b_1$) $(a_{s-1},b'_2)$, meaning that $b'_2=b_{r}$ when $b'_2$
exists, and $b_1=b_{r}$ when $b'_2$ does not exist. We have two (sub)cases:

\begin{enumerate} 
\item[B.1.]  When $b'_2$ exists, we show that the sequence 
$U^1=(a_1,b_1)\ldots, (a_{s-1},b_1), (a_s,b_1), (a_s,b'_2)$ is the sought canonical sequence.
Clearly, $(a_{s-1},b_1)$ is obtained from $(a_1,b_1)$ using the setting sequence $U^0$
from which $(a_{s-1}, b_r)$ is useless in this case. Also, $(a_s, b'_2)$ is obtained from
$(a_s, b_1)$ and $(b_1,b'_2)$ by transitivity (case 3 in Claim \ref{claim:imminduce}).
It remains to show that $(a_s,b_1)$ is deduced from $(a_{s-1},b_1)$ and $\{a_{s-1}, a_s\}$.
In $U$, $\{a_{s-1}, a_s\}$ is used to deduce $(a_s,b_r)$ from $(a_{s-1},b_{r})$,
using either case 1 or case 2 in Claim \ref{claim:imminduce}. If case 1 is used, then
$(a_s,a_{s-1})$ is an arc (new or old), and it allows to deduce $(a_s,b_1)$ from 
$(a_{s-1},b_{1})$ using the transitivity. If case 2 is used, then $\{a_{s-1}, a_s\}$
is a local basis, thus $b_1$ is a top of it. The resulting NB-constraint allows to deduce
$(a_s,b_1)$ from  $(a_{s-1},b_{1})$ in this case too. 
\item[B.2] When $b'_2$ does not exist, we have that $b_r=b_1$ and $U^1:=U^0.(a_s,b_1)$
is a canonical setting sequence for $(a_s,b_r)$.

\end{enumerate}

\bfn
Let $F$ be a directed linear profile on  $[n]\cup\{0,n+1\}$. Then 
the NB-transitive closure $G^1$ obtained in step  5 of Algorithm \ref{algo:linear} 
when $b_1$ (respectively $a_1$) are chosen as in step 3 (respectively step 4) has no circuit.
\label{claim:circuit}
\efn

{\bf Proof.} Assume a circuit $d_1,d_2, \ldots, d_c$, $c\geq 2$, is created in $G^1$. Because of
the transitive closure, a shortest such circuit has length 2. Let then $d_1,d_2$ form
a $2$-circuit and
assume that (at least) $(d_1,d_2)$ is a new arc. Then, according to Claim \ref{claim:shortpath},
there exists a canonical setting path with vertices $a_1, \ldots, a_s(=d_1)$ and 
$b_1, \ldots, b_r (=d_2)$  ($r\in\{1,2\}$).  
Consequently $(d_2,d_1)$ cannot be an old arc, since then
in $G$ either we have directly that $(b_1,d_1)$ is an old arc (when $r=1$ and thus $d_2=d_1)$
or the transitivity guarantees the same conclusion when $r=2$. But this  yields  a contradiction 
with Claim \ref{claim:before5}(a).

We deduce that $(d_2,d_1)$ is a new arc, implying again the existence of a canonical setting path
with vertices $a'_1, \ldots, a'_{s'}(=d_2)$ and $b'_1, \ldots, b'_{r'} (=d_1)$  ($r'\in\{1,2\}$).
But $b'_1=b_1$ and $a'_1=a_1$. Consequently  we have either that $b_1=d_1$ (when $r'=1$) or
 that $(b_1,d_1)$ is an old arc (when $r'=2$). In the former case we have a contradiction
with affirmation $(a)$ in the definition of a canonical setting path since $d_1=a_s=b_1$. In the latter case, we have again a contradiction with Claim \ref{claim:before5}(a). $\Box$
\bigskip

We are now ready to prove the main theorem:
 
 \begin{thm}
 {\sc Directed $\M$-Betweenness} is polynomial for linear $\M$-profiles.
 \end{thm}

{\bf Proof.} Given a linear $\M$-profile $F$, we first apply Algorithm \ref{algo:Arcs} and,
if it returns a pair $(G,\Sil)$, we apply Algorithm \ref{algo:linear}. To  show the correctness
of the algorithm, we show the answer is "No" iff there is no permutation whose $\M$-profile is $F$.

If the answer is "No", then Algorithm \ref{algo:Arcs} returns that $G$ is not a DAG, which occurs
iff some $\M$-constraints from $F$ cannot be simultaneously satisfied. Thus, there is no permutation 
with $\M$-profile $F$.

Now, assume there is no permutation whose $\M$-profile is $F$, and suppose by contradiction that
the algorithm returns a permutation $P$. We show that $P$ satisfies all the $\M$-constraints in $F$, 
yielding a contradiction with the hypothesis. The permutation $P$ is output at the end of Algorithm \ref{algo:linear},
showing that Algorithm \ref{algo:Arcs} finishes with a DAG $G$. 
Then, in
Algorithm \ref{algo:linear} every execution of the {\bf while} loop in steps 1-6 satisfies at least
one silent NB-constraint and, by Claim \ref{claim:circuit}, creates no circuit. Therefore, 
the pair $(G,\Sil)$ obtained at the end of each execution consists again in a DAG $G$ 
with $R-$, $B-$, $T$- and $NB$-arcs, and a set $\Sil$ with smaller size than the previous one.
Thus the {\bf while} loop ends when $\Sil=\emptyset$ and yields a DAG $G$ that satisfies
all the constraints imposed by the $\M$-profile $F$. Any topological order of $G$, including $P$, is thus
a permutation with $\M$-profile $F$. The hypothesis that there is no permutation with
$\M$-profile $F$ is thus contradicted.

The execution time of the algorithm is clearly dominated by the $|\Sil|$ computations of
the NB-transitivity closure in step 5 of Algorithm \ref{algo:linear}. Now, the number
of NB-constraints in $\Sil$ is in $O(n^2)$ (we have at most one NB-constraint $\neg(t\lrf{a}t+1)$ for each
$t$ and each $a$) and the NB-transitivity closure is clearly performed in polynomial time,
thus the resulting algorithm is polynomial.  $\Box$

\section{Generalizations}\label{sect:Generalizations}

In this section, we generalize the definition of $\M$-profiles so as to allow
them to carry different amounts of information, depending on an integer parameter $k$.

\bdefin
Let $k$ be a positive integer with $1\leq k\leq n+1$. The {\em $\kM$-profile of a permutation} $P$ on $\n\cup\{0,n+1\}$ 
is the set of {\em $\kM$-constraints}

$$M_k(P)=\bigcup_{i=1}^{k}\{t \frac{\,\, \scriptstyle{[\min_{t,t+i}, \max_{t,t+i}]}}{}\,\, t+i\,\, |\,\, 0\leq t\leq n+1-i\}$$

\noindent where $\min_{t,t+i}$ ($\max_{t,t+i}$ respectively) is the minimum (maximum respectively) value 
in the interval delimited on $P$ by the element $t$ (included) and the element $t+i$ (included). 
\edefin

Note that $\M$-profiles as defined in Section \ref{sect:def} are the $\oM$-profiles.
A $\kM$-profile is thus a $\M$-profile augmented with longer-range information of the same
nature as the $\M$-profile itself,  for pairs $\{t,t+i\}$ with $i$ at most equal to $k$.
Then all the definitions  related to $\M$-profiles generalize to $\kM$-profiles in an obvious way,
allowing us to state the following variant of the $\M$-Betweenness Problem:

\bigskip

\noindent {\sc (directed or not) $\kM$-$\M$  Betweenness}

\noindent{\bf Input:} A positive integer $n$, a (directed or not) $\kM$-profile $F_k$ on $\n\cup \{0, n+1\}$. 

\noindent{\bf Question:} Is there a permutation $P$ on $\n\cup\{0,n+1\}$ whose $\kM$-profile is $F_k$? 
\bigskip

Similarly to the $\M$-{\sc Betweenness} problem, the  $\kM$-$\M$ {\sc Betweenness} problem provides a $\kM$-profile,
which imposes B-constraints and NB-constraints for the permutations associated with it, if any. 
The existence of at least one permutation raises the question of its uniqueness,
allowing to know whether the permutation is characterized by its $\kM$-profile or not.
More formally, we state the two following problems:

\bigskip

\noindent {\sc (Directed or not) $\M$-Reconstruction}

\noindent{\bf Input:} A positive integer $n$.

\noindent{\bf Requires:} Find the minimum value of $k$ such that (directed or not) $\kM$-$\M$ {\sc Betweenness} 
has at most one solution, for each possible $\kM$-profile $F_k$ on $\n\cup\{0,n+1\}$.
\bigskip

\noindent {\sc (Directed or not) Unique $\kM$-$\M$ Betweenness}

\noindent{\bf Input:} A positive integer $n$, a (directed or not) $\kM$-profile $F_k$ on $\n\cup \{0, n+1\}$. 

\noindent{\bf Requires:} Decide whether $F_k$ is the $\kM$-profile of a unique permutation on $\n\cup\{0,n+1\}$,
or not. In the positive case, find the unique permutation associated with $F_k$.
\bigskip

Problems $\kM$-$\M$ {\sc Betweenness} and  {\sc Unique $\kM$-$\M$ Betweenness} are clearly related,
but do not allow easy  deductions in one sense or the other. For instance, even if we have
a solution for the {\sc Directed $\M$-Betweenness} in the case of a linear profile 
(see Section \ref{sect:Particular}), we know nothing about the uniqueness of the permutation $P$
the algorithm outputs (when such a permutation exists).

%\rt{Deviation de la permutation "$\M$-Identite" qui serait celle obtenue d'un profile
%en demandant que tout entre $m_t$ et $M_t$ soit entre $t$ et $t+1$ ? Attention, il se
%pourrait que l'identite n'existe pas ....}

In the subsequent, we solve the $\M$-{\sc Reconstruction} problem in the undirected case,
and give a lower bound for the directed case.  We assume wlog that the 
$\kM$-profiles we use are compatible with the assumption that 0 and $n+1$ are respectively 
the leftmost and the rightmost element in the permutations we are dealing with. 
Then we prove the following result:

\begin{thm}
The minimum $k$ in (directed or not) $\M$-{\sc Reconstruction} satisfies:

\begin{enumerate}
\item[$(a)$]  $k=\max\{1,n-3\}$ in  $\M$-{\sc Reconstruction}.
\item[$(b)$] $k\geq  \lceil \frac{n-3}{2}\rceil$ in {\sc Directed $\M$-Reconstruction}, for $n\geq 4$. For $n=1, 2, 3$, we have $k=1$.
\end{enumerate}
\label{thm:bounds}
\end{thm}

The proof is based on the following claim. 
\begin{fait}
Let $n$ be a positive integer, and $F_k$ be a (directed or not) $\kM$-profile on $\n\cup\{0,n+1\}$ $(1\leq k\leq n+1)$. Then: 

\begin{enumerate}
\item In all the permutations whose $\kM$-profile is $F_k$ (if any),
the elements $1$ and $n$ have precisely the same positions, denoted $q_1$ and $q_n$
\item If $l=\min\{q_1,q_n\}$ and $r=\max\{q_1,q_n\}$, then the sets $X, Y$ and $Z$ of elements 
situated  respectively strictly between the positions 0 and $l$ (for $X$), $l$ and $r$ (for $Y$), 
$r$ and $n+1$ (for $Z$) are the same over all the permutations with $\kM$-profile $F_k$ (if any).
\end{enumerate} 
\label{claim:1n}
\end{fait} 

\noindent{\bf Proof.} Assuming at least one permutation corresponding to $F_k$ exists, let $P$ be such
a permutation. Denote $q_1$ the position of $1$ on $P$ and successively consider the B-constraints

$$n\lrf{m_{n,n+1}}n+1, n-1\lrf{m_{n-1,n}}n, \ldots, 2\lrf{m_{2,3}}3.$$ 

\noindent The first B-constraint places
$n$ on the left of $1$ iff $m_{n,n+1}=1$, the second one places $n-1$ on the opposite side of $1$ with respect to
$n$ iff $m_{n-1,1}=1$ and so on. Each element in $\{n, n-1,\ldots, 2\}$ is deterministically placed on the
left or on the right of $1$ depending only on those B-constraints.  As a consequence, $1$ is at the same place 
$q_1$ in all permutations corresponding to $F_k$.

A similar reasoning may be done with the element $n$ and the B-constraints: 

$$0\lrf{M_{0,1}}1, 1\lrf{M_{1,2}}2, \ldots, n-2\lrf{M_{n-2,n-1}}n-1.$$ 

\noindent We similarly deduce that $n$ is at the same place $q_n$ in all permutations corresponding to $F_k$,
and the sets of elements situated respectively on its left and right are the same in all permutations.

Putting together the previous deductions, whatever the order of $q_1$ and $q_n$,
we have that - on the one hand - $X\cup Y$ and $Z$ are identical in all permutations, and
- on the other hand - $X$ and $Y\cup Z$ are identical in all permutations. The conclusion follows. $\Box$
\bigskip

{\bf Proof of Theorem \ref{thm:bounds}.}  We now prove affirmations $(a)$ and $(b)$.

\noindent{\sl Proof of affirmation $(a)$.} For $n=1$, it is trivial.
When $n\in\{1, 2, 3\}$ it is easy to prove,  using Claim \ref{claim:1n}, that the $1$-profile guarantees 
the uniqueness of the associated permutation.
When $n\geq 4$, assume by contradiction that  $k<n-3$ and  let $P$ be a permutation on $\n\cup\{0,n+1\}$ 
whose elements in positions 1 to 4 are $p_1=k+2$, $p_2=k+3$, $p_3=1$ and $p_4=n$.
Let $F_k$ be the $\kM$-profile of $P$.
According to Claim  \ref{claim:1n}, the elements $1$ and $n$ are situated respectively at positions
3 and 4 in all permutations associated with $F_k$, and positions 1 and 2 are occupied (whatever the order) 
by the elements $k+2$ and $k+3$. Now, in $F_k$ the 
$\kM$-constraints involving one of the elements $k+2$ and $k+3$  and another
element following $1$ on its right are useless for fixing the places of $k+2$ and $k+3$ 
since these constraints have the minimum and maximum element $1$ and $n$.
The only possibly useful $\kM$-constraints are those involving $0, 1, k+2$ and $k+3$, but these 
integers have pairwise difference larger than $k$ except for $k+2$ and $k+3$. Now,  $k+2$ and $k+3$ are 
involved in the $1$-constraint $k+2\frac{\scriptstyle{[k+2,k+3]}}{}k+3$, which does not fix them on the places 1 and 2 of
the permutation. Thus, there are at least two permutations with $\kM$-profile $F_k$, a contradiction. 
We thus have $k\geq n-3$.
 
We now show that if $k=n-3$, then there is at most one permutation on $\n$ whose $\kM$-profile is 
$F_k$. This is shown by induction on $n$. 

When $n=4$ and $k=1$, Claim \ref{claim:1n} guarantees that, if at least one permutation with the
given $1$-profile exists, then $1$ and $4$ have fixed places, and 
$2$ (respectively $3$) is located in the  same set among  $X, Y, Z$ in all suitable permutations.
If $2$ and $3$ are in different sets, then the uniqueness is guaranteed. Otherwise, either $2$ and $3$ are in a set delimited
by the position of $1$, and then the constraint $1\frac{\,\scriptstyle{[m_{1,2},M_{1,2}]}\,}{}2$ allows
to deduce whether $3$ separates $1$ and $2$ or not (thus fixing the positions of 2 and 3), 
or they are in a set delimited by $n(=4)$, and then the
constraint $3\frac{\,\scriptstyle{[m_{3,4},M_{3,4}]}\,}{}4$ allows to deduce whether $2$ separates $3$ and $4$ or not.
In all cases, all the elements are located at fixed places, thus the permutation associated with the
$1$-profile is unique.

Assume now, by inductive hypothesis, that for all $n'< n$,  a $(n'-3)$-profile either has no
associated permutation, or has exactly one. Let now $F_{n-3}$ be a $(n-3)$-profile for permutations on 
$\n\cup\{0,n+1\}$,  and let $q_1, q_n, l,r, X, Y, Z$ be defined according to  Claim \ref{claim:1n}, assuming at least one
permutation exists. Denote $P$ any of these permutations, extended with $0$ and $n+1$. 
Let $W=X\cup\{0,n\}$, if $q_n<q_1$,  and $W=X\cup Y\cup \{0, 1, n\}$, otherwise. We show
that:

\begin{equation}
\mbox{The elements in}\, W\, \mbox{have fixed places in any permutation}\ P 
\label{eq:three}
\end{equation}

Note that $P[0..q_n]$, whose element set is $W$,  is a subpermutation of $P$ delimited by $0$ and $n$,  
which are respectively the  
minimum and maximum element in $P[0..q_n]$. Now, renumber the elements of  $P[0..q_n]$ from $0$ 
to $n'+1$ according to their increasing values, where $n'<n$ and $n'+1$ is at position $q_n$. 
Then the resulting permutation is a permutation $P'$ on $[n']$ augmented with $0$ and $n'+1$.

Denote  $F_{n'-3}$ the $(n'-3)$-profile of this permutation, and let us show  that $P'$ is unique.
For $n'\geq 4$, we show that when $F_{n-3}$ is known, 
$F_{n'-3}$ is also known, and then apply inductive hypothesis  to deduce that $F_{n'-3}$ (and thus $F_{n-3}$) fixes 
the places of the elements in $P[0..q_n]$.   Whereas for $n'=2, 3$ 
we show that there are enough 1-constraints deduced from   $F_{n-3}$ to guarantee the uniqueness of $P'$.
The case $n'=1$ is trivial.

Let $h\frac{\,\scriptstyle{[m_{h,h+i},M_{h,h+i}]}\,}{}h+i$ be a constraint on $P'$, which belongs to $F_{n'-3}$ if $n'\geq 4$
({\em i.e.} $1\leq i\leq n'-3$) and to $F_1$ if $n'=2,3$.   Let $b(h), b(m_{h,h+i}), b(M_{h,h+i})$ and $b(h+i)$ 
be respectively the labels of $h, m_{h,h+i},M_{h,h+i}, h+i$ before renumbering. Then the difference between the labels
of $h$ and $h+i$ in the initial $P$ satisfies:

\begin{equation}
b(h+i)-b(h)\leq (h+i)-h+(n-n'-1).
\label{eq:this}
\end{equation}

\noindent Indeed, if $x$ elements of $P$ are between $n$ and $n+1$, then the total number of elements in $P$
is, on the one hand, $n+2$ (the cardinality of $\n\cup\{0,n+1\}$) and, on the other hand, $1+n'+1+x+1$
(given by  the cardinality of $W$, by $x$ and by the element $n+1$,). Then $x=n-n'-1$, and it represents
the maximum number of elements that can miss between $b(h+i)$ and $b(h)$, additionally to the values
separating them in $P'$, {\em i.e.} $(h+i)-h$. But then from equation (\ref{eq:this}) we deduce: 

\begin{equation}
b(h+i)-b(h)\leq i+n-n'-1.
\label{eq:bh}
\end{equation}

{\sl Case $n'\geq 4$.} From equation (\ref{eq:bh}) we deduce with $i\leq n'-3$ that $b(h+i)-b(h)\leq n'-3+n-n'-1\leq n-4$,
meaning that $b(h)\frac{\,\scriptstyle{[b(m_{h,h+i}),b(M_{h,h+i})]}\,}{}b(h+i)$ is
a constraint from $F_{n-3}$, yielding the constraint $h\frac{\,\scriptstyle{[m_{h,h+i},M_{h,h+i}]}\,}{}h+i$
of $F_{n'-3}$ after renumbering. Of course, this affirmation is true since the renumbering
keeps the order between the elements, and thus the (renumbered) minimum and maximum value of each
given interval. Thus $F_{n'-3}$ is deducible from $F_{n-3}$ and, by inductive hypothesis, the
permutation $P'$ is uniquely determined by $F_{n'-3}$.

{\sl Case $n'=2$.} Then, as assumed above, $i=1$ and thus  equation (\ref{eq:bh}) implies 
$b(h+1)-b(h)\leq 1+n-2-1=n-2$ which is larger than $n-3$. This shows that all the $1$-constraints
on $P'$ with $b(h+1)-b(h)\leq n-3$ are deducible from constraints in $F_{n-3}$ but the 
$1$-constraints on $P'$ with $b(h+1)-b(h)=n-2$ are not. These latter $1$-constraints 
are obtained when $b(h+1)-b(h)= 1+n-2-1$ (according to equation (\ref{eq:bh})), that is, when
$b(h+1)-b(h)= n-2$. To achieve this with $0, n$ and the other two elements $e,f$ in $W$ (w.l.o.g. assume
$e\leq f$) we must have either $e=n-2$, and thus $f=n-1$, (such that $e-0=n-2$), or $e=1$ and $f=n-1$ (such that
$f-e=n-2$), or $e=1$ and $f=2$ (such that $n-f=n-2$). In all cases, exactly one $1$-constraint is missing
({\em i.e.} not resulting from $F_{n-3}$) but the uniqueness of the permutation is still guaranteed,
since the two other $1$-constraints are sufficient to fix the elements in a $4$-permutation (including
the endpoints $0$ and $4$).

{\sl Case $n'=3$.} Using (\ref{eq:bh}) and the information that $i=1$, we deduce that 
$b(h+1)-b(h)\leq 1+n-3-1=n-3$, and thus all the $1$-constraints are available for $P'$.
As the theorem is true for permutations on $3$ elements, then we are done.

Affirmation (\ref{eq:three}) is proved.
Similarly, we show that the places of the elements situated on each permutation $P$ between the element $1$
(in position $q_1$) and the element $n+1$ are fixed. Thus, all the elements of each permutation $P$ are  in
fixed places, and there is only one permutation $P$ with the $(n-3)$-profile $F_{n-3}$.  
\medskip

\noindent{\sl Proof of affirmation $(b)$.} Similarly to the undirected case, in the directed case assume by contradiction that 
$k<\lceil \frac{n-3}{2}\rceil$ and build $P$ as in the
undirected case, but with $p_2=2k+3$ instead of $p_2=k+3$ (thus avoiding  the directed
$\kM$-constraint $k+2\lf{[k+2,k+3]}k+3$ or  $k+2\rf{[k+2,k+3]}k+3$, which fixes in the directed case
the positions of  $k+2$ and $k+3$). 
Then, with the directed $\kM$-profile $F_k$ of $P$, no $\kM$-constraint exists involving $0, 1, k+2, 2k+3$, 
and thus $k+2$ and $2k+3$ may permute on the positions $1$ and $2$. Therefore, at least two permutations exist
with the $\kM$-profile $F_k$, a contradiction.

Thus the uniqueness of the permutation implies $k\geq \lceil \frac{n-3}{2}\rceil$. 
$\Box$

\section{Conclusions and Perspectives}\label{sect:Conclusion}

In this paper, we investigated some problems related to the construction of
a permutation from a $\M$-profile or, more generally, from some $\kM$-profile,
with $1\leq k\leq n$. For the first of these problems, the $\M$-{\sc Betweenness} problem,
we noticed the main difficulties of the directed version and gave a polynomial particular case.

The undirected version is even more difficult, due to differences with respect to the
directed version that we  present hereafter. First, as the relative position of $t$ and $t+1$
({\em i.e.} the arc of $G$ between $t$ and $t+1$) is not directly given by the 
$\M$-profile, the B-constraints cannot be directly exploited as in steps 3-5 of 
Algorithm \ref{algo:Arcs}. The construction of those two types of arcs, the $R$-arcs and
the $B$-arcs, must therefore be integrated into the Build-Closure algorithm, 
where the B-constraints must be considered as well as the NB-constraints when seeking  
new arcs to be added to $G$. It may be noticed that, similarly to the case of the NB-constraints,  
any of the B-constraints $t\lrf{m_t}t+1, t\lrf{M_t} t+1$ has two possible settings, resulting either in the  
set of new  arcs 
$Arcs^+=\{(t,t+1), (t,m_t), (t,M_t), (m_t,t+1), (M_t,t+1)\}$, or in the set of new arcs
$Arcs^-=\{(t+1,t), (m_t,t), (M_t,t), (t+1, m_t), (t+1,M_t)\}$. When one arc is set, then
the four other arcs are set accordingly. When no arc is set, the B-constraints are silent.
The algorithm obtained from Algorithm \ref{algo:Arcs} by performing the indicated changes
thus outputs either the answer No, or $G$ and two  sets $\Sil$ and $SilB$ of
silent NB- and silent B-constraints respectively.
We thus arrive at the second main difference between the directed and undirected case. 
Any setting sequence must allow to deduce new arcs also using the B-constraints, 
thus adding  cases to those already in Claim \ref{claim:imminduce}, and yielding the
study of the arc propagation in $G$ even more complicated that in the directed case.

For both versions, and also for the more general $\kM$-$\M$ {\sc Betweenness} problem, the 
algorithmic difficulty of the problem is an open problem. The same holds for the
{\sc Directed $\M$-Reconstruction} problem. Also, being able to recognize
a $\kM$-profile allowing to reconstruct exactly one permutation, {\em i.e.} solving
(Directed or not) {\sc Unique} $\kM$-$\M$ {\sc Betweenness}, would allow to
identify a subclass of permutations perfectly represented by their $\kM$-profile.

\bibliographystyle{plain}
\bibliography{ProfPerm}
\end{document}